\documentclass[prl,twocolumn,aps,10pt]{revtex4-1}

\usepackage{graphicx}  
\graphicspath{{./Figures/}}
\usepackage{dcolumn}  
\usepackage{amssymb, amsmath}
\usepackage{hyperref}

\begin{document}

\title{Portable magnetometry for detection of biomagnetism in ambient environments}
\author{M.\ E.\ Limes}
\author{E.\ L.\ Foley}
\author{T.\ W.\ Kornack}
\affiliation{Twinleaf LLC, 300 Deer Creek Dr., Plainsboro, New Jersey, 08536, USA}
\author{S. Caliga}
\author{S. McBride}
\author{A. Braun}
\affiliation{SRI International, 201 Washington Rd., Princeton, New Jersey, 08540, USA}
\author{W.\ Lee}
\author{V. G.\ Lucivero}
\author{M.\ V.\ Romalis}
\affiliation{Department of Physics, Princeton University, Princeton, New Jersey, 08544, USA}

\date{\today}
\begin{abstract}
We present a method of optical magnetometry with parts-per-billion resolution that is able to detect biomagnetic signals generated from the human brain and heart in Earth's ambient environment.
Our magnetically silent sensors measure the total magnetic field by detecting the free-precession frequency of highly spin-polarized alkali metal vapor. 
A first-order gradiometer is formed from two magnetometers that are separated by a 3 cm baseline.
Our gradiometer operates from a laptop consuming 5 W over a USB port, enabled by state-of-the-art micro-fabricated alkali vapor cells, advanced thermal insulation, custom electronics, and laser packages within the sensor head.  The gradiometer obtains a sensitivity of 16 fT/cm/Hz$^{1/2}$ outdoors, which we use to detect neuronal electrical currents and magnetic cardiography signals. Recording of neuronal magnetic fields is one of a few available methods for non-invasive functional brain imaging that usually requires extensive magnetic shielding and other infractructure. This work demonstrates the possibility of a dense array of portable biomagnetic sensors that are deployable in a variety of natural environments. 

\end{abstract}
\maketitle
%
Magnetoenephalography (MEG) and electroencephalography (EEG) serve as important windows into human brain function by providing neuronal current source imaging with millisecond resolution---much faster than other  noninvasive  techniques, such as functional magnetic resonance imaging (fMRI) and positron emission tomography (PET) \cite{Hamalainen_1993}. MEG also has several advantages over EEG, including improved source localization and non-contact  measurements \cite{Baillet_2017}. 
Commercially available MEG systems use superconducting quantum interference device (SQUID)  magnetometers or gradiometers with  sensitivity of 3-10 fT/Hz$^{1/2}$. However, existing MEG systems require large, expensive magnetically shielded rooms or human-scale magnetic shields, as well as dewars and infrastructure for cryogenic operation. Placing subjects in a magnetically shielded room restricts the range of behaviors and activities that can potentially be studied.  There are a few demonstrations of MEG detection using SQUIDs in an unshielded environment \cite{Vrba_1995,Okada_2006,Seki_2010}, but such recordings rely on third-order gradiometers that are primarily sensitive to shallow neuronal current sources.  SQUID gradiometers are fundamentally limited in the cancellation of uniform magnetic fields by the fabrication tolerances of their pick-up coils. 

As an alternative to the cyrogenically cooled and bulky SQUID MEG systems, there has been a recent surge in research of optically pumped magnetometers for MEG detection. Most sensitive atomic magnetoemters operate using alkali vapors  near zero field in a spin-exchange relaxation free (SERF) regime \cite{Kominis_2003}. SERF magnetometers have been used for detection \cite{Xia_2006,Johnson_2010,Sander_2012,Johnson_2013,Shah_2013,Alem_2014,Sheng_2017} and localization \cite{Kim_2014, Boto_2017,Boto_2018,Borna_2020} of MEG signals, but still require magnetic shielding or field cancellation because their operation relies on having a small total magnetic field. In addition, they require individual calibration and have inherently limited linearity and dynamic range. All measurements with wearable SERF magnetometers have been performed so far in magnetically shielded rooms \cite{Boto_2018, Hill_2019}.

We present a method of operating a portable optical gradiometric sensor that is able to detect MEG signals in Earth's ambient magnetic field, all while exposed to natural magnetic noise sources. Unlike all previous MEG sensors, our technique uses two total-field magnetometers that directly measure the Larmor precession frequency of alkali vapor electron spins in the magnetic field. Frequency measurements have a much greater dynamic range and linearity compared to voltage measurements associated with other magnetic field sensors. Furthermore, they do not require individual calibration, so we simply subtract the frequencies recorded from two alkali vapor cells to find a first-order magnetic field gradient. A first-order gradiometer in principle allows for detection of deeper current sources. We demonstrate the performance of our sensor by detecting MEG in Earth's ambient environment, as well detecting human heartbeats in real time in magnetically noisy environments.
\begin{figure*}
\includegraphics[width=\textwidth]{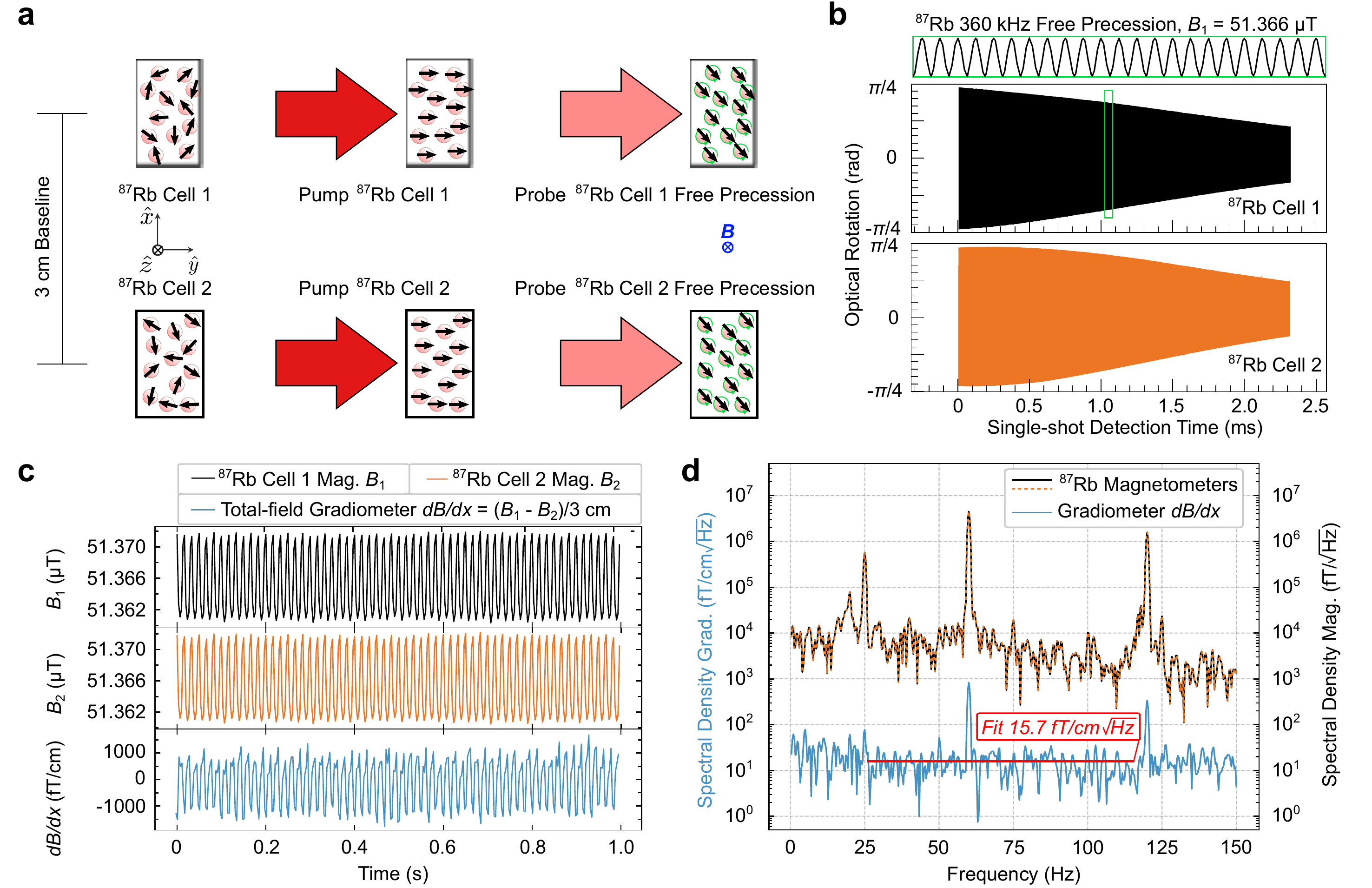}
\caption{{\em Magnetic gradiometer operation.} a) Two $^{87}$Rb vapor cells separated by a 3 cm baseline are optically pumped with a laser pulse train to near unity polarization.  A quantum non-demolition measurement of the $^{87}$Rb polarization along the probe axis occurs with a weak detuned laser undergoing optical rotation. 
$^{87}$Rb undergoes free precession about a total field $B$, leading to a decaying sine wave signal for each magnetometer cell. The total field $B$, and quantization axis, is essentially determined by the local Earth's field $B_E$. 
b) A custom frequency counter extracts a $^{87}$Rb free-precession frequency for each cell by detecting optical-rotation zero crossings in a single-shot detection period of 2.3 ms. The counter then calculates a total field for each cell, $B_1$ and $B_2$, by dividing by the $^{87}$Rb gyromagnetic ratio $\gamma/2 \pi \approx 7$ kHz/$\mu$T. 
c) The single-shot pump and probe measurement is repeated at a rate of 300 Hz. The total fields $B_1$ and $B_2$ are streamed to a laptop, and subtracted to obtain the first-order total-field gradient $dB/dx$.
d) We take the spectral densities of the total-field magnetometers and gradiometer time-domain data to demonstrate the unshielded noise floor and common-mode rejection of the all-optical gradiometer in Earth's ambient environment. The red line is a fit to the gradiometer noise floor giving 15.7 fT/cm$\sqrt{\text{Hz}}$.  }
\label{fig:1}
\end{figure*}

{\em Portable biomagnetic sensor.} Our portable gradiometer uses two $8 \times 8 \times 12.5$ mm$^3$ $^{87}$Rb vapor cells separated by 3 cm. These cells have anodically bonded glass windows with internal mirrors for 795 nm light \cite{Dural_2017}. 
The cells are evacuated, baked, and filled with enriched $^{87}$Rb and 650 torr N$_2$. In operation they are electrically heated inside of a radiation shield that has low magnetic noise \cite{Dang_2010}. 
The cells are attached to a glass substrate with low-thermal-conduction supports and are placed into a $6.5 \times 1.8 \times 1.8$ cm$^3$ cuvette that is evacuated and sealed to maintain high vacuum, eliminating gas conduction and convection. This assembly requires 30 mW/cell to heat to the operating temperature of 100$^{\circ}$C, which gives a $^{87}$Rb density of about $5\times 10^{12}$ atoms/cm$^3$. The vacuum packaging and low-power laser modules allow result in the outside surface temperature of sensor to be 32$^{\circ}$C, below body temperature.

Shown in Fig.~\ref{fig:1}a, we form a magnetic gradiometer by orienting the two alkali vapor cells such that multi-pass laser beams optically pump and probe $^{87}$Rb in a plane transverse to Earth's field $B_\text{E} \approx 51.4$ $\mu$T. 
A multi-pass pump is used due to geometrical constraints imposed by the radiation shielding. Thus the high intensity pump pulses travel along the path of the probe beam, which limits the dead-zone to one dead axis. 
The total fields each vapor cell experiences, $B_1$ and $B_2$, are dominated by Earth's field $B_\text{E}$, and the direction of $B_\text{E}$ essentially determines the quantization axis.
$^{87}$Rb atoms are spin-polarized to near unity by an on-resonant (D1) 795 nm pulsed pump diode laser that is able to produce several W for $\mu$s pulses. 
The beam is sent through a polarizer and $\lambda/4$ wave plate to make $\sigma^+$ light for optical pumping. 
We use a style of optical pumping similar to a conventional Bell-Bloom scheme \cite{Bell_1961, Gerginov_2017}, obtaining a high initial atomic spin polarization by pumping synchronously with a pump pulse train at the Larmor precession rate active only during the state initialization period.
$^{87}$Rb atomic spins are pumped in the transverse plane to near unity polarization, which causes a suppression of the dominant spin-exchange relaxation mechanism between the $F=2$ and $F=1$ hyperfine manifolds at these sizable magnetic fields, and leads to an extension in $^{87}$Rb coherence time $T_2$ \cite{Romalis_2016,Lucivero_20}.  
 We then stop pumping and detect the magnetic field during a $^{87}$Rb free-precession period in order to eliminate frequency shifts  associated with the pump laser.
A 0.1 mW linearly polarized VCSEL probe beam is far-detuned from resonance (D1) and undergoes paramagnetic Faraday rotation in a multi-pass configuration that yields high signal-to-noise \cite{Li_2011, Sheng_2013,Lee_20}. 
The probes of each cell are sent into balanced polarimeters that measure signals corresponding to $^{87}$Rb free precession about the total field, shown in Fig.~\ref{fig:1}b.  After roughly 1 ms of state initialization and dead time, our acquisition time is 2.3 ms per shot. The entirety of the optics and lasers are housed in a 3D-printed case along with a photodiode amplifier (PDA) board.  
\begin{figure*}
\includegraphics[width = \textwidth]{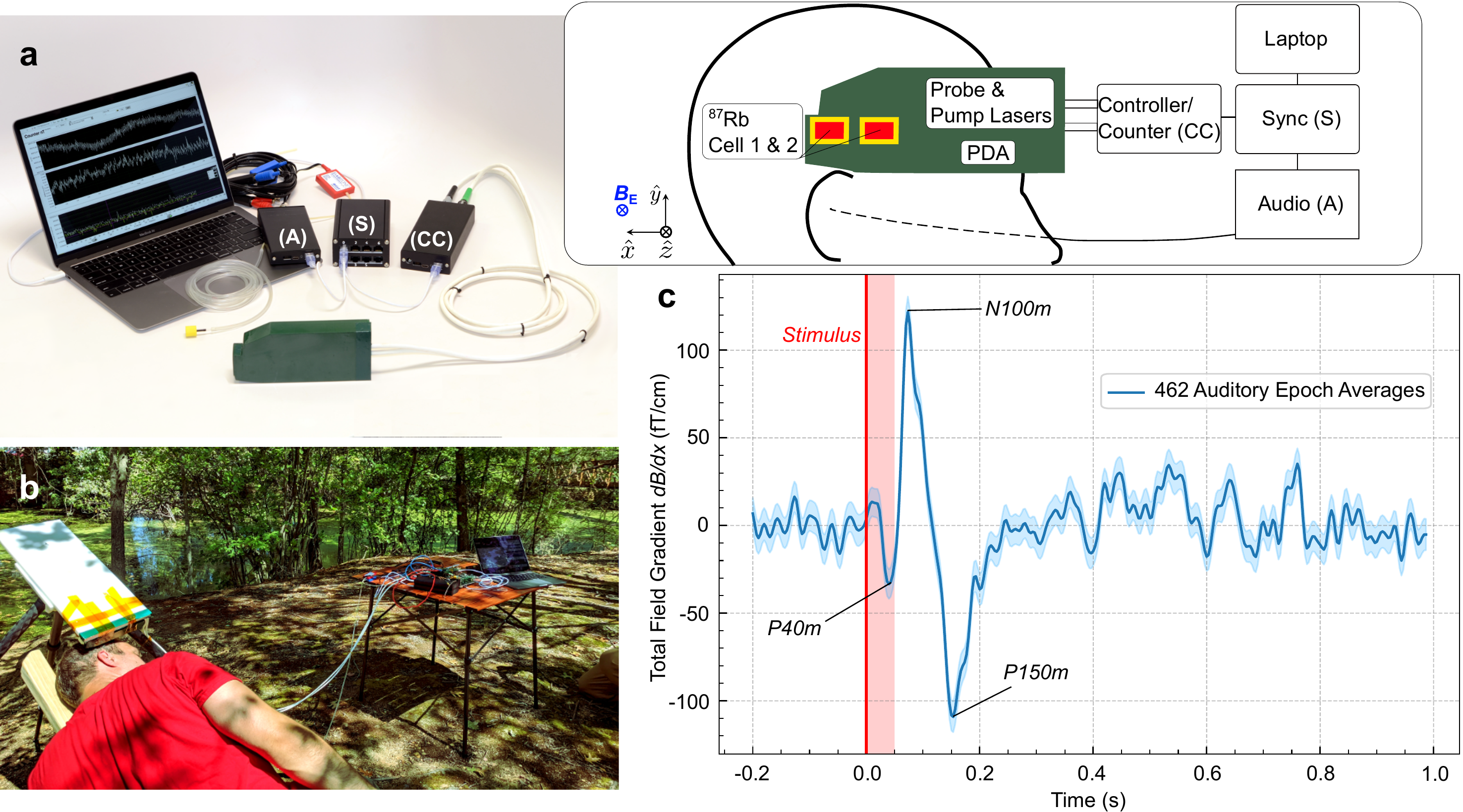}
\caption{{\em Auditory evoked fields detected unshielded in Earth's field with a portable first-order gradiometer.} a) The vacuum-packaged $^{87}$Rb vapor cells are optically pumped and probed by diode lasers integrated into the gradiometer head. Optical rotation of the probe laser is detected with a photodiode amplifier assembly (PDA). The gradiometer is controlled by compact electronics, including a custom counter streams the total fields $B_1$ and $B_2$ and gradient $dB/dx$ to the laptop. The two $^{87}$Rb vapor cells are placed near a subject's auditory cortex to measure the total-field gradient $dB/dx$ with a 3 cm baseline. 
b) A picture of the in-the-field MEG recording apparatus with a subject.
c) For a given subject, data is recorded for roughly 20 min while time-randomized 1 kHz auditory stimuli are applied to the left ear of the subject. A filtered average of 462 epochs is shown for $dB/dx$ gradient data taken above a subject's right ear.  Bands indicate the standard error of the mean. The auditory evoked fields observed include the prominent N100m peak along with indication of P40m and P150m responses. 
}
\label{fig:2}
\end{figure*}

The electronics for the gradiometer consist of two $6.4 \times 12.7$ cm$^2$ PCB boards each powered by 2.5 W from a  USB laptop port (or a 5V battery). One board controls the sensor heating and probe laser, while the other contains a frequency counter and controls the pump-probe sequence. We use 50 kHz for driving the heater, which we find to not add any additional noise to the measurement. The frequency counter streams two time-stamped frequencies for each acquisition period to a Labview VI for real-time analysis and logging. The power consumption of the boards is dominated by the microcontrollers and can be reduced with  available lower-power versions.  We use a shot-to-shot repetition rate of 300 Hz that also determines that data rate of the fields of both magnetometers streamed to the laptop. 
Subtracting the two measured fields $B_1$ and $B_2$ and dividing by the 3 cm baseline we find the total field gradient $dB/dx$. An example of data received by the computer is shown in Fig.~\ref{fig:1}c. 

 To demonstrate the system's potential for biomagnetic measurements, we show the spectral densities of the streamed magnetometer and gradiometer data in Fig.~\ref{fig:1}d with a human subject's head near the sensor, in a natural environment. The largest peaks observed by the sensor are 60 and 120 Hz components coming from power lines roughly 75 m away. Another prominent peak at 25 Hz we find comes from a nearby New Jersey Transit/Amtrak Rail line about 750 m away. Considering the 3 cm baseline, the 60 Hz peak suppression of the gradiometer indicates a common mode rejection ratio of at least 2000.  We obtain a similar noise level measured without a subject, and we note that it is difficult to separate real gradient noise from the intrinsic noise floor of the gradiometer in an unshielded environment without multiple gradiometers. Ignoring the 60 Hz peak in a least-squares fit, the gradiometer spectral density between 26 and 115 Hz gives 15.7 fT/cm$\sqrt{\text{Hz}}$; this implies a magnetometer noise floor of 33.3 fT/$\sqrt{\text{Hz}}$, which outperforms commercially available scalar atomic sensors operating unshielded in Earth's field.
Within magnetic shielding, the total-field gradiometer achieves 10 fT/cm$\sqrt{\text{Hz}}$  with a field of 50 $\mu$T applied.

{\em Ambient MEG.} Operating in a new regime enabled by the first-order gradiometer, we demonstrate detection of MEG signals in Earth's ambient environment, choosing to focus on auditory evoked field responses. Here, 1 kHz audio stimuli of duration 50 ms is generated with a delivery time randomized in a $2.5\pm1$~s window to prevent subject adaptation.  These stimuli are delivered by a non-magnetic pneumatic earphone that is placed into an ear of the subject. The gradiometer is held by a non-magnetic mount near the audio cortex above the opposing ear, as the subject rests in a wooden chair. The center of each vacuum packed alkali vapor cell is about 1.2 cm away from the head of the subject which is smaller than the typical distance for SQUID magnetometers in conventional MEG systems.  This measurement stand-off is determined by the mean distance of the optical path to cell edge and is increased from radiation shielding and mounting of the cuvette to the sensor case. A synchronization board is used to ensure the precise time-stamping of the audio stimuli referenced to the data streamed from portable gradiometer system, as shown in Fig.~\ref{fig:2}a. The gradiometer is oriented such that Earth's field $B_\text{E}$ is aligned to optimize the signal for a current dipole expected from auditory evoked responses, with an example of the subject's orientation with respect to the sensor in Fig.~\ref{fig:2}b. 
Auditory evoked field data was recorded for four subjects in several 5-20 min trials, using different sensor positions and orientation. 


We show in Fig.~\ref{fig:2}c MEG data that is filtered and averaged over 462 epochs for a subject in a particular trial. 
For MEG data analysis we apply a 0.5-50 Hz bandpass filter, along with 25, 60, and 120 Hz notch filters, and observe P40m, N100m, and P150m evoked fields (the relatively fast N100m response is consistent with contralateral stimulation for a relatively short interstimulus interval \cite{Makela_1993,Gutschalk_2019}). 
 Auditory evoked signals were detected in all four subjects and the sign of the detected $dB/dx$ peaks is consistent with the orientation of the current dipole observed in previous studies of auditory evoked fields. 
The orientation of the sensor and head with respect to the Earth's field is critical to MEG operation, as total-field magnetometers are only sensitive to the component of the biomagnetic field parallel to the bias field. 
Field changes in the transverse plane $B_\text{T}$  appear only in second order $B_\text{T}^2/2B_\text{E}$ and are negligible.
Thus when considering arrays of this type of total-field sensor, additional signal processing will be required for making constraints on source localization methods.

\begin{figure}
\includegraphics[width=3.33in]{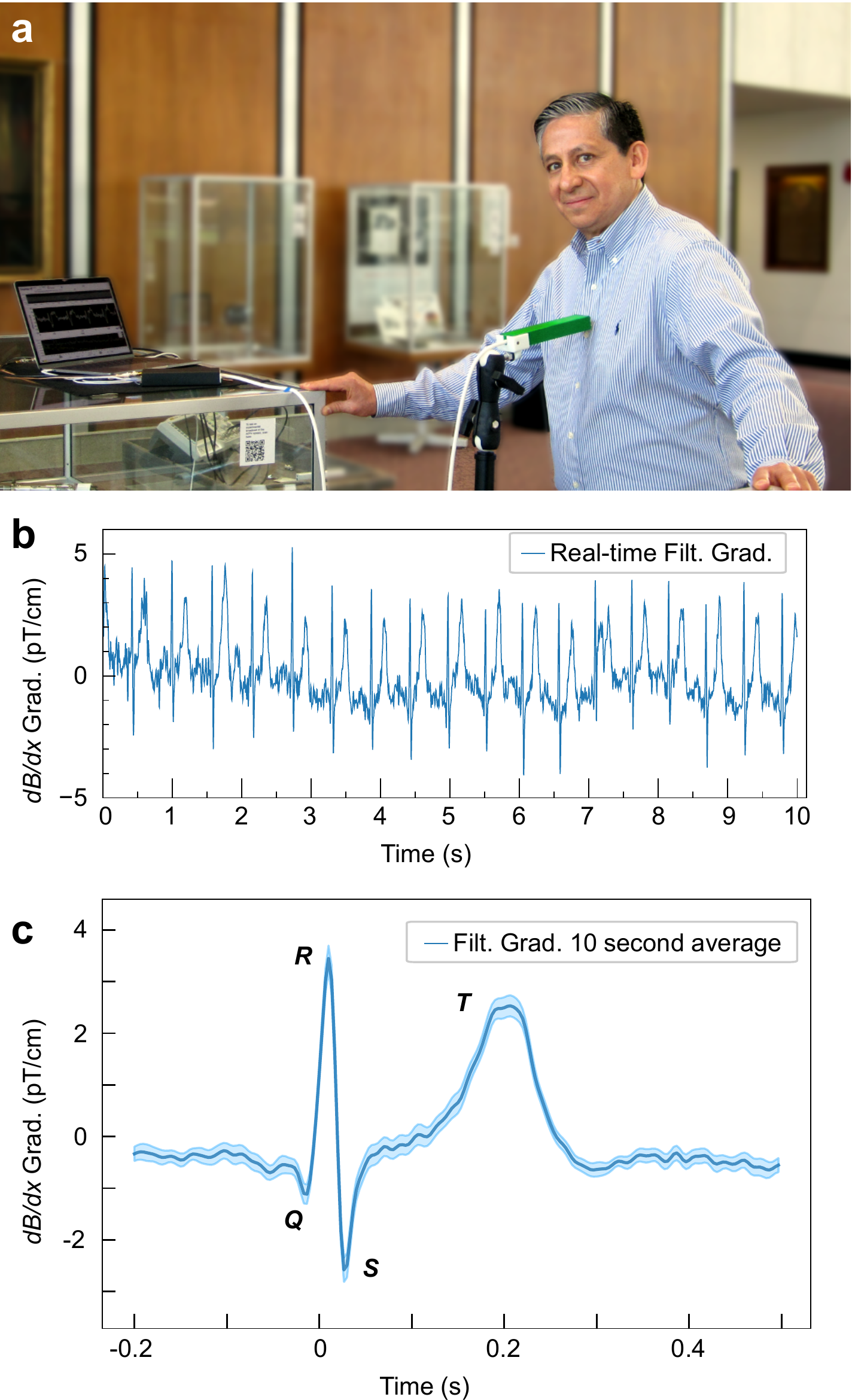}
\caption{ {\em Field-deployable magnetocardiography.} a) A picture of a next-generation gradiometer that is able to detect human heartbeats in magnetically noisy environments. b) As a subject presents their chest to the sensor, a real-time filter (bandpass 0.1-50 Hz notch at 60/120 Hz) is applied to the total field gradient $dB/dx$ to demonstrate heartbeat measurements. c) The result of averaging data for 10 s, triggered on the R peak.}
\label{fig:HB}
\end{figure}

 We briefly mention that another important medical application of sensitive magnetometers is magnetocardiography (MCG). Magnetic fields generated by the heart are stronger, but their detection in magnetically unshielded, noisy environments typical of a research laboratory or hosptial remains a challenge. A number of optically-pumped sensors are being developed for this application \cite{Bison_2003,Wyllie_2012,Alem_2015,Bai_2019}. 
In Fig.~\ref{fig:HB} we show real-time in-the-field MCG signals that are taken by simply walking up and presenting the chest to the sensor, along with a 10 s averaging of human heartbeats. 
Even in its current form our sensor is not far from providing a practical MCG device that allows quick, electrode-free heart diagnostics for triage in ambient environments. 

With this work, we have presented a method of operating optical magnetometers in Earth's natural environment with unprecedented performance. We showed the potential of this technique for biomagnetic measurements through proof-of-principle detections of unshielded MEG and MCG signals using a portable first-order gradiometer. Wearable atomic sensors that do not require shielding will enable a greater variety of MEG research studies, as well as reduce their cost. They can  eventually replace EEG sensors currently being used in a variety of  open-source EEG systems \cite{openBCI}. By using an all-optical design we have eliminated cross-talk between sensors, which is a deficiency of sensors that require RF or microwave fields. This important feature allows for these types of sensors to be formed into a scalable array, which is necessary for proper source localization, as well as to suppress magnetic gradient noise from power lines and other nearby sources, and enables practical operation in a typical laboratory or hospital environment. We also note that the sensitivity of the gradiometer can be further improved \cite{Sheng_2013} to compete with SQUID sensitivities, while retaining the ability to measure closer to the biomagnetic source than SQUID systems.  We believe the development of these types of portable sensors for ambulatory subjects, that can be used in a cost-effective scalable array, will impact the scope of many MEG and MCG research studies, as well as the host of various other applications that can benefit from a commercially available sensor for total-field  magnetometry in Earth's natural environment.  

Sensor development was funded by the Defense Advanced Research Projects Agency (DARPA) Microsystems Technology Office (MTO) under Contract No. 140D6318C0020. The views, opinions and/or findings expressed are those of the authors and should not be interpreted as representing the official views or policies of the Department of Defense or the U.S. Government. Approved for Public Release, Distribution Unlimited.
Proof-of-principle demonstrations of the sensor detecting biomagnetic signals was supported by Princeton University and the Fetzer Franklin Fund of the John E. Fetzer Memorial Trust.”
\linebreak

%

\onecolumngrid

\newpage

\section{Supplementary Information}

\setcounter{figure}{0}   
\begin{figure}[h!]
\includegraphics[width=3.5in]{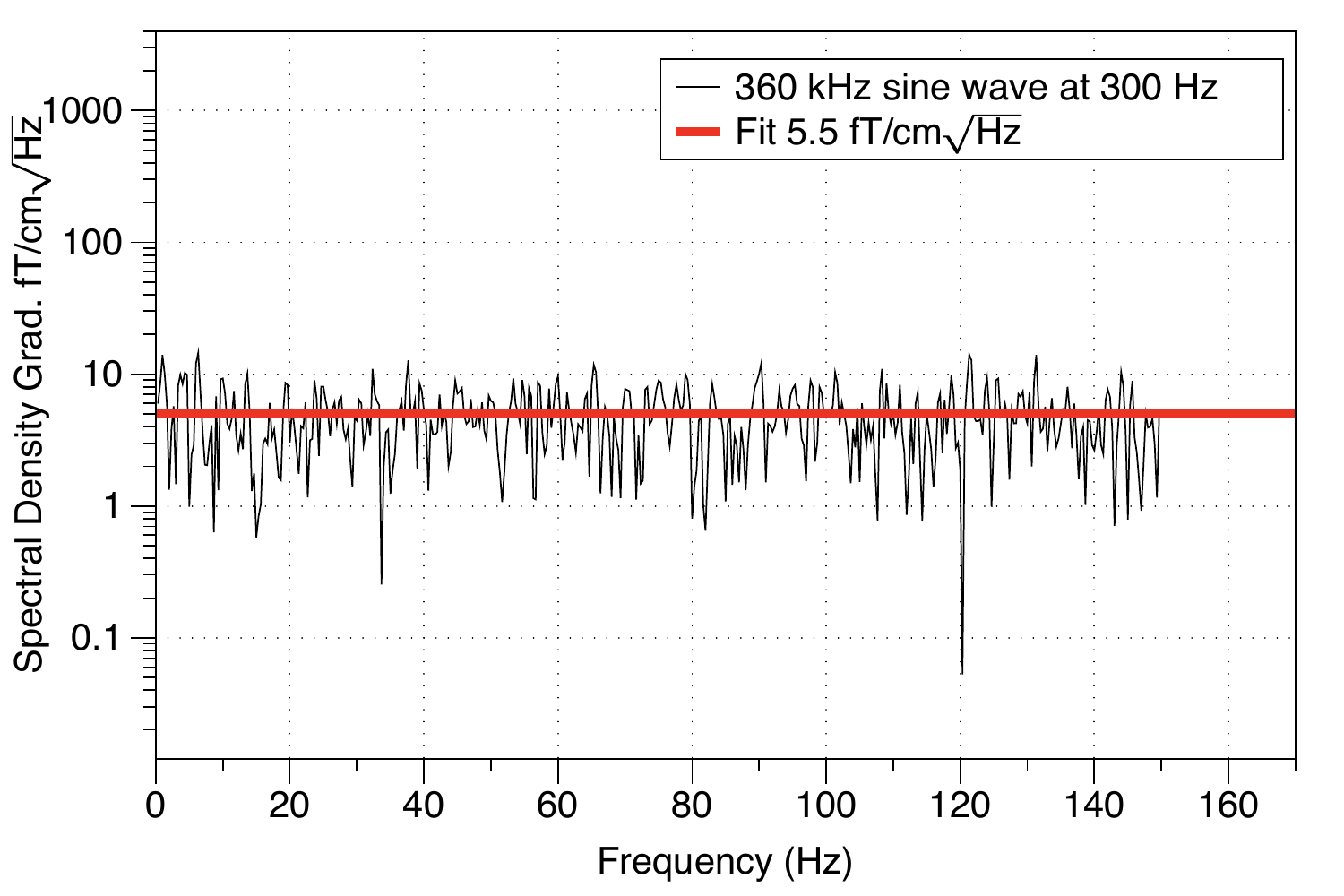}
\caption{ The custom frequency counter is set up under the operating conditions used in the main text, triggered on 300 Hz and acquisition window of 2.3 ms. 360 kHz sine waves from a function generator are sent into the two channels. A fit to the white noise gives our reported portable detection noise floor of 5.5 fT/cm$\sqrt{\text{Hz}}$.
}
\label{fig:bm}
\end{figure}

\begin{figure}
\includegraphics[width=3.5in]{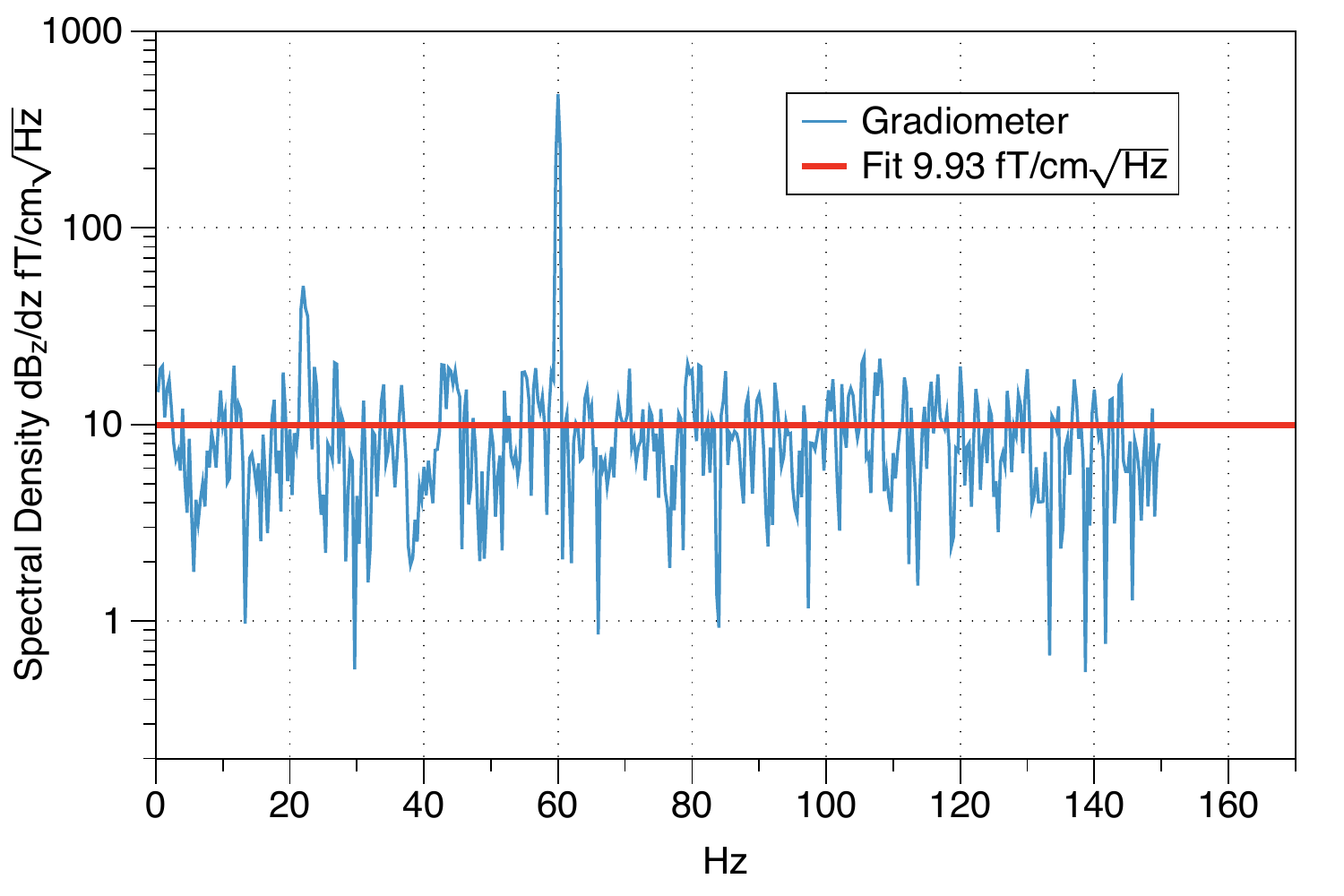}
\caption{Spectral density of gradiometer in magnetic shielding (Twinleaf MS-2), taken with 49.3 $\mu$T bias field. }
\label{fig:Gen2Noise}
\end{figure}

\begin{figure}[h!]
\includegraphics[width=3.5in]{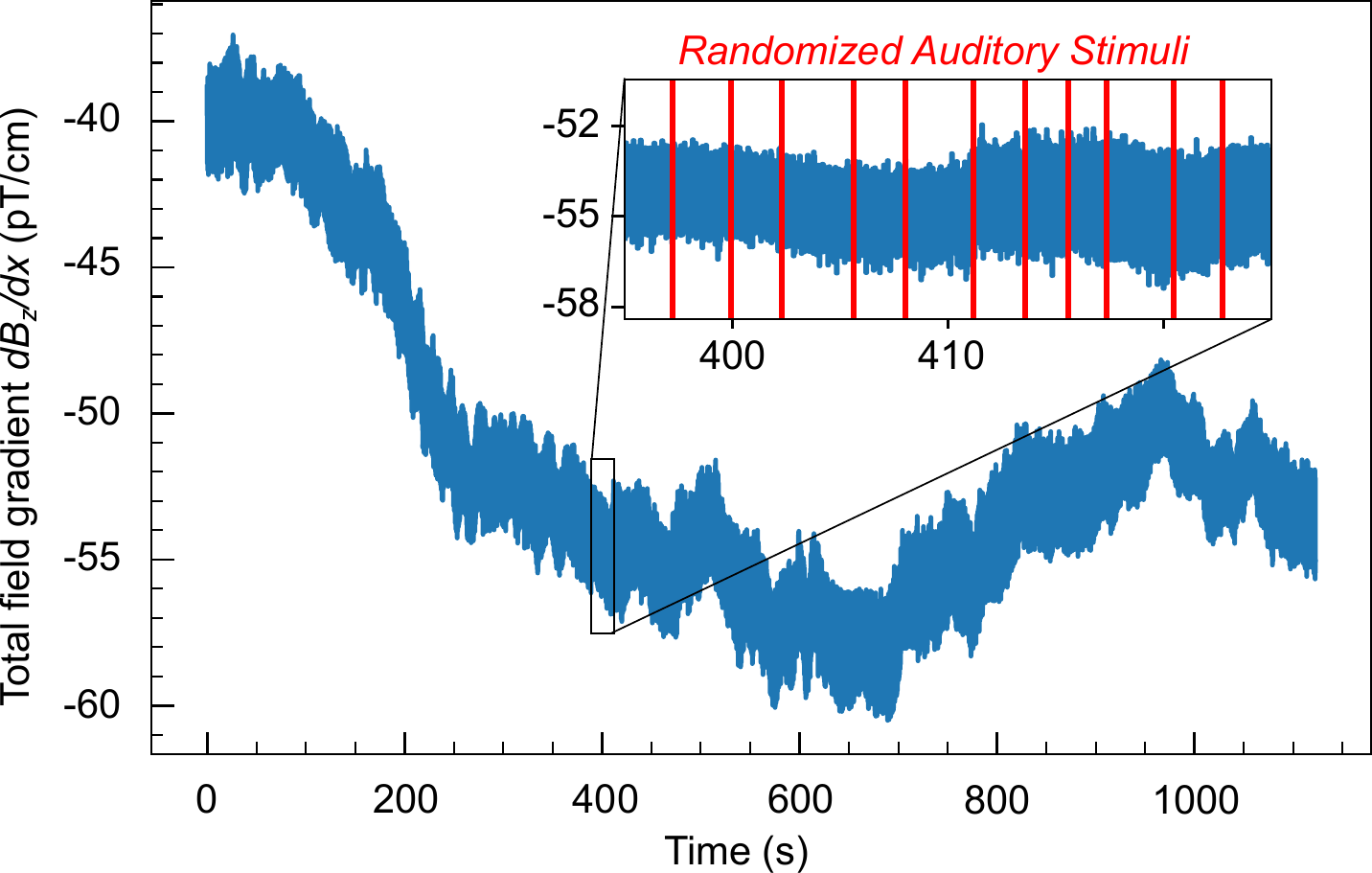}
\caption{ The raw $dB_z/dx$ gradient data shown averaged in Fig. 3 of the main text. }
\label{fig:raw}
\end{figure}
\begin{figure}
\includegraphics[width=7in]{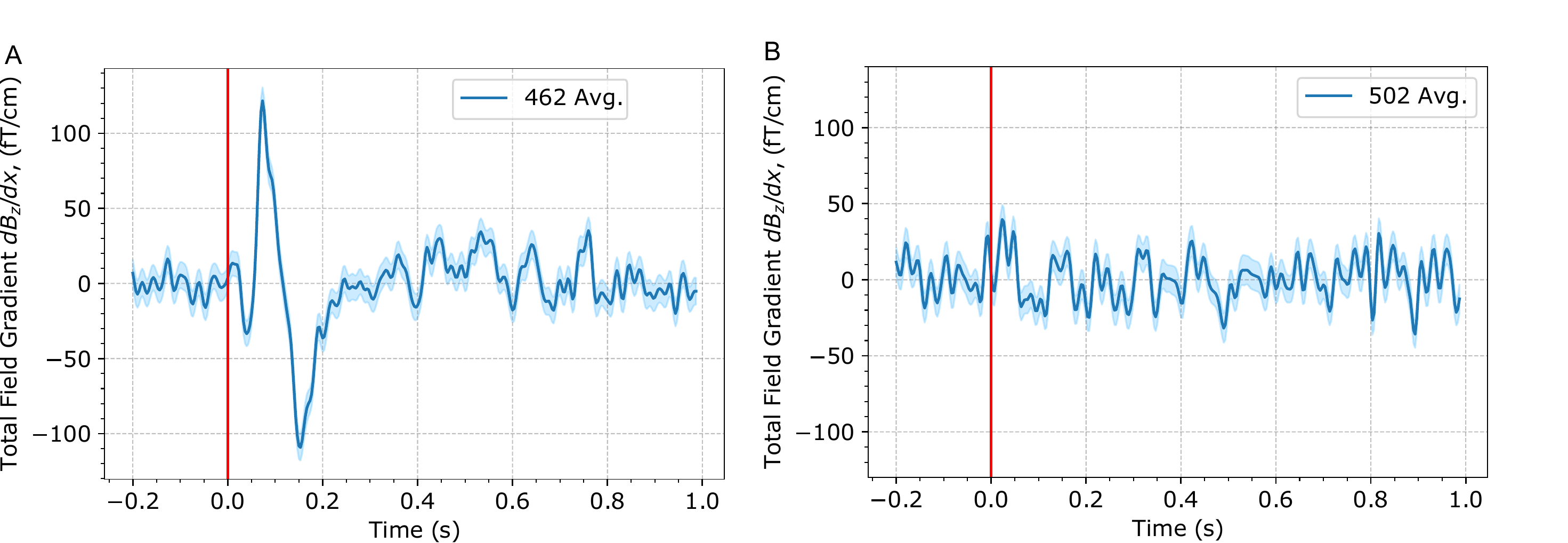}
\caption{Data for subject with (A) and without (B) auditory stimuli. }
\label{fig:onoff}
\end{figure}

\begin{figure}[h!]
\includegraphics[width=7in]{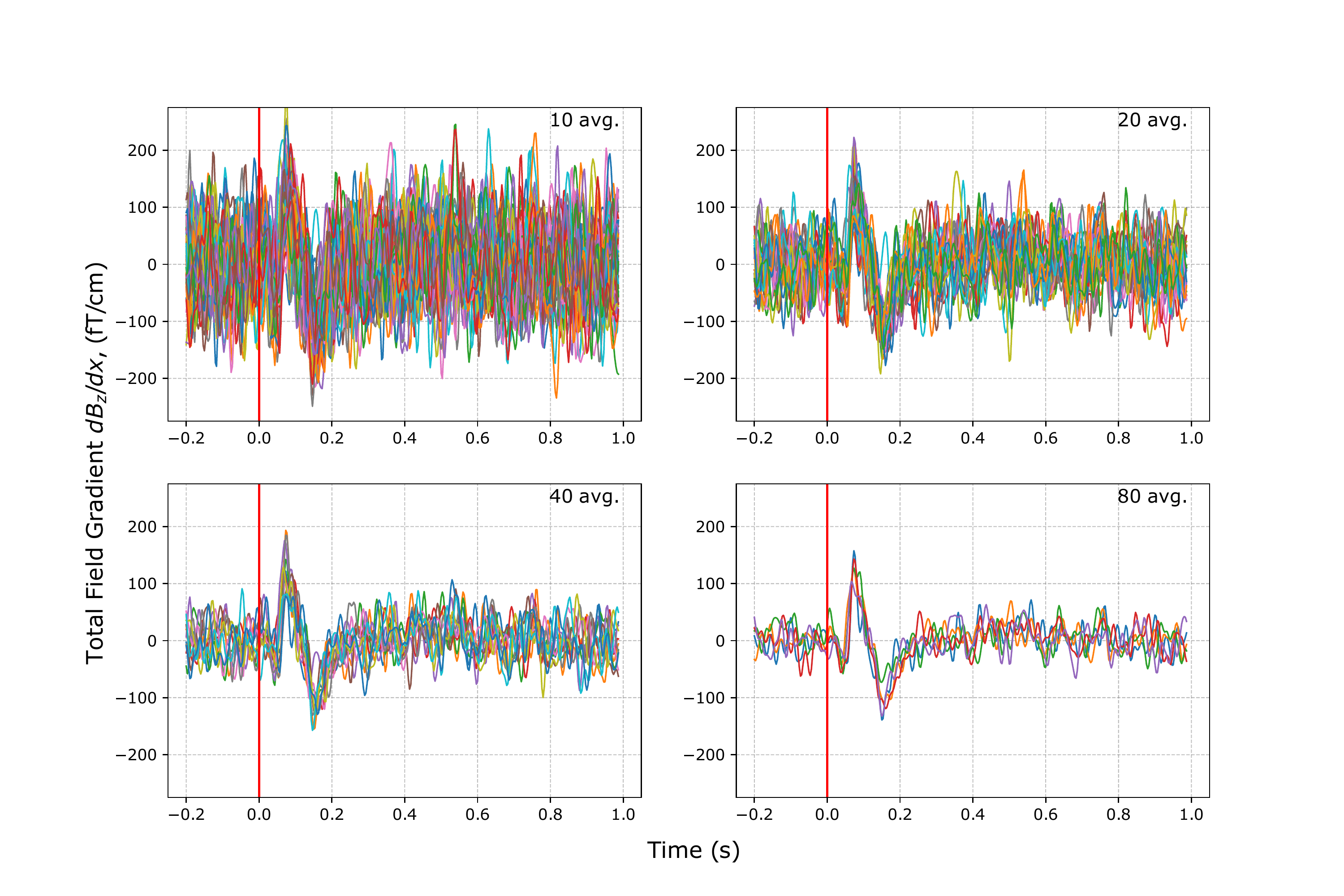}
\caption{ Taking averages (10, 20, 40, and 80) of the data taken shown in Fig.~3 of the main text.  }
\label{fig:avg}
\end{figure}`

\begin{figure}[h!]
\includegraphics[width=7in]{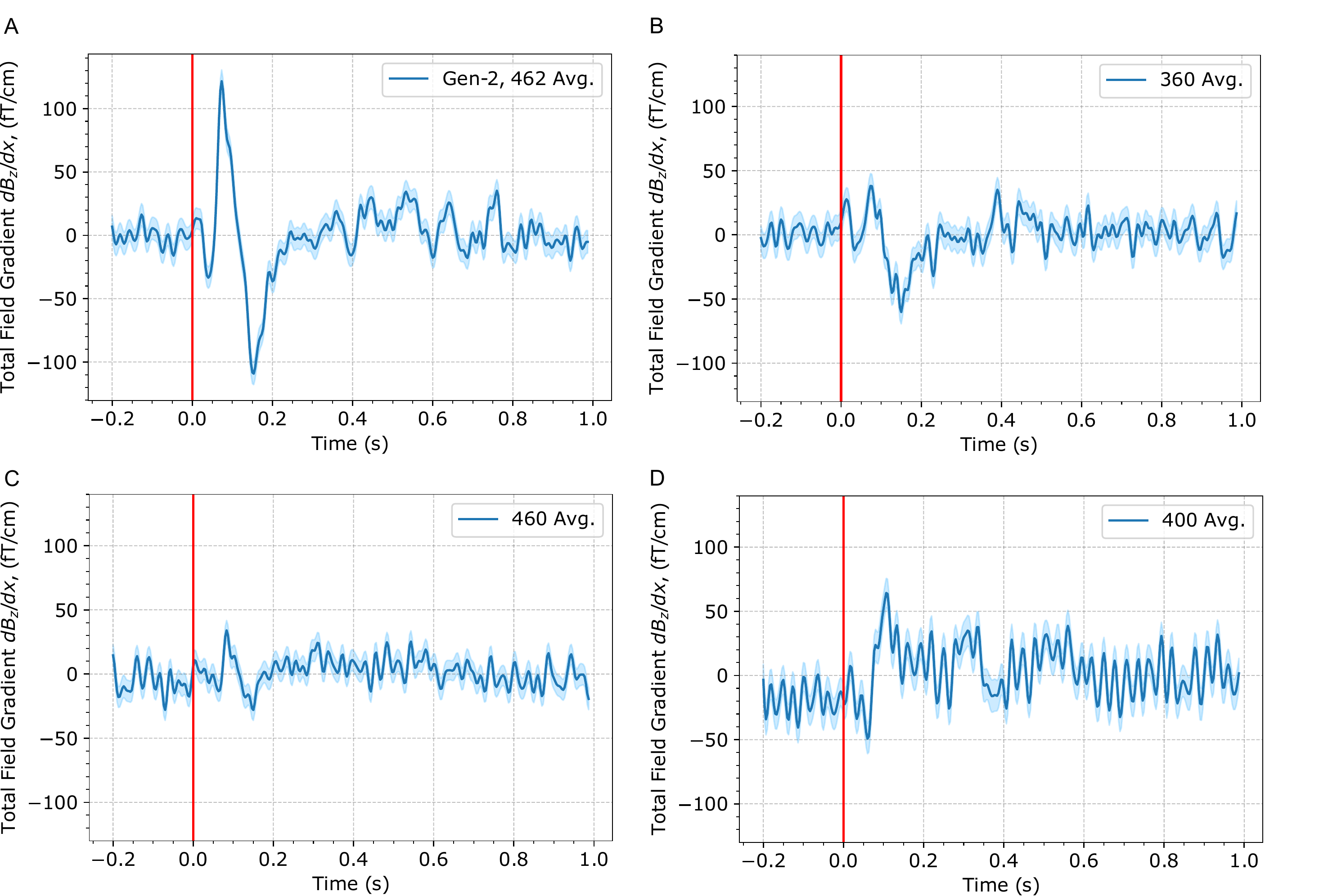}
\caption{ Data taken for four different subjects showing auditory evoked fields.  }
\label{fig:foursubs}
\end{figure}`

\end{document}